\documentclass [
  sigconf,
  screen,
  nonacm
] {acmart}

\usepackage[T1]{fontenc}
\usepackage{amsmath}
\usepackage{multirow}
\usepackage{hyperref}
\usepackage[capitalize]{cleveref}
\usepackage{listings}
\usepackage{framed}
\usepackage{tabularx}
\usepackage{booktabs}
\usepackage{footmisc}
\usepackage{algorithmic}
\usepackage{graphicx}
\usepackage{textcomp}
\usepackage{xcolor}
\usepackage{url}
\usepackage{relsize}
\usepackage{microtype}
\def\BibTeX{{\rm B\kern-.05em{\sc i\kern-.025em b}\kern-.08em
		T\kern-.1667em\lower.7ex\hbox{E}\kern-.125emX}}

\crefname {section}    {Sec.}    {sections}
\Crefname {section}    {Section} {Sections}
\crefname {table}      {Tbl.}    {tables}
\Crefname {table}      {Table}   {Tables}
\crefname {lstlisting} {Lst.}    {listings}
\Crefname {lstlisting} {Listing} {Listings}

\newcommand \parhead [1]{\vspace{+.8mm}\noindent \textbf {{#1}}}

\newcommand{\ghrepo}[2]{%
	\href{#2}{\texttt{\textcolor{blue}{#1}}}%
}

\usepackage[normalem]{ulem}

\usepackage{ifthen}
\newboolean{showcomments}
\setboolean{showcomments}{true} 
\ifthenelse{\boolean{showcomments}}
  {\newcommand{\nb}[2]{
    \fcolorbox{gray}{yellow}{\bfseries\sffamily\scriptsize#1}
    {\sf\small\textcolor{teal}{$\blacktriangleright$\textit{#2}$\blacktriangleleft$}}
  }
  
  }
  {\newcommand{\nb}[2]{}
  
  }

\AtBeginDocument{%
  \providecommand\BibTeX{{%
    Bib\TeX}}}
\setcopyright{acmlicensed}
\copyrightyear{2018}
\acmYear{2018}
\acmDOI{XXXXXXX.XXXXXXX}

\lstdefinestyle{cppstyle}{
	language=C++,
	basicstyle=\ttfamily\small,
	keywordstyle=\bfseries\color{blue},
	commentstyle=\itshape\color{green},
	stringstyle=\color{red},
	numberstyle=\tiny\color{gray},
	stepnumber=1,
	numbersep=10pt,
	backgroundcolor=\color{white},
	showspaces=false,
	showstringspaces=false,
	showtabs=false,
	frame=single,
	tabsize=2,
	captionpos=b,
	breaklines=true,
	breakatwhitespace=true,
	escapeinside={(*@}{@*)}, 
	morekeywords={nullptr, size_t, uint8_t}
}
\begin{document}

\title{Beyond the Control Equations: An Artifact Study of Implementation Quality in Robot Control Software}

\author{Nils Chur}
\orcid{0009-0009-2427-8342}
\affiliation{%
  \institution{Ruhr University Bochum}
  \city{Bochum}
  \country{Germany}
}
\email{nils.chur@rub.de}

\author{Thorsten Berger}
\affiliation{%
  \institution{Ruhr University Bochum}
  \city{Bochum}
  \country{Germany}}
\email{thorsten.berger@rub.de}
\orcid{0000-0002-3870-5167}

\author{Einar Broch Johnsen}
\affiliation{%
 \institution{University of Oslo}
 \city{Oslo}
 \country{Norway}}
\email{einarj@ifi.uio.no}
\orcid{0000-0001-5382-3949}

\author{Andrzej Wąsowski}
\affiliation{%
	\institution{IT University of Copenhagen}
	\city{Copenhagen}
	\country{Denmark}}
\email{wasowski@itu.dk}
\orcid{0000-0003-0532-2685}




\begin{abstract}\looseness=-1
A controller---a software module managing hardware behavior---is a key component of a typical robot system.  While control theory gives safety guarantees for standard controller designs, the practical implementation of controllers in software introduces complexities that are often overlooked. Controllers are often designed in continuous space, while the software is executed in discrete space, undermining some of the theoretical guarantees. Despite extensive research on control theory and control modeling, little attention has been paid to the implementations of controllers and how their theoretical guarantees are ensured in real-world software systems.

	We investigate 184 real-world controller implementations in open-source robot software.  We examine their application context, the implementation characteristics, and the testing methods employed to ensure correctness. We find that the implementations often handle discretization in an ad hoc manner, leading to potential issues with real-time reliability.  Challenges such as timing inconsistencies, lack of proper error handling, and inadequate consideration of real-time constraints further complicate matters.  Testing practices are superficial, no systematic verification of theoretical guarantees is used, leaving possible inconsistencies between expected and actual behavior. Our findings highlight the need for improved implementation guidelines and rigorous verification techniques to ensure the reliability and safety of robotic controllers in practice.

\end{abstract}





\maketitle


\section{Introduction}%

Robots are increasingly used in safety critical domains, including transportation, healthcare, and manufacturing\,\cite{GUIOCHET201743}. Robotic systems are characterized by the tight coupling of software and physical hardware. A key component in robot systems is the \emph{controller}, which generates the signals to manage hardware actuation behavior, ensuring essential properties, such as stability and robustness\,\cite{YU201546}. Controllers thus bridge the gap between software and the physical environment. Implemented in software, they govern real-world system behavior and form the staple of safety and reliability.

\begin{figure} [t]
\vspace{+.3cm}
	\includegraphics [
    width = \linewidth,
    trim = .5mm .5mm .5mm .5mm,
    clip
  ]{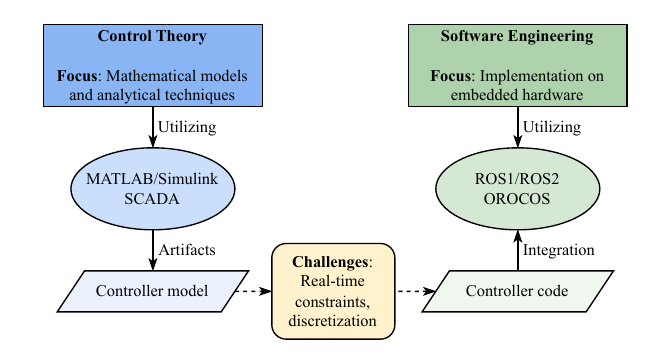}

  \vspace{-2mm plus 0.5mm minus 0.5mm}

  \caption{The different perspectives of control and software engineers and the associated challenges.}%
  \label{fig:GapCT-SE}
\vspace{-.1cm}
\end{figure}

While control theory does provide a well-established foundation for designing controllers with guarantees of stability, robustness, and response time\,\cite{glad2018control, bubnicki2005modern, 6413422}, significantly less attention has been given to the implementation of controllers in software, and whether the guarantees from control theory actually hold in practice in the controller code.
Controllers in robotic systems operate in a discrete software environment, despite having been designed in the continuous domain.
Discretization---the transformation of continuous control algorithms into discrete implementations---is essential to maintain stability and robustness, but its challenges are often underestimated.
Control engineering does offer a broad theory of controller discretization, but how these principles are reflected, or neglected, in real controller implementations remains unclear\,\cite{ogata1995discrete}.
\Cref{fig:GapCT-SE} summarizes the different perspectives
on controllers and associated challenges, as seen from control theory and from software engineering.
In practice, controller behavior can be affected by sampling rates, execution timing, and jitter\,\cite{cervin2003does}. These factors can cause deviations between the theoretical controller design and its actual runtime behavior, potentially undermining guarantees provided by the theory\,\cite{ogata1995discrete, di2019sampled}.

When implementing controllers, developers
need to consider software engineering challenges, including
modularity, maintainability, and integration into complex robotic
systems. 
Middleware, such as ROS\,\cite{ROS1,ROS2}, OROCOS\,\cite{1242011}, and YARP\,\cite{Metta:YARP:2006}, facilitates the integration of robotic software components (such as perception, planning, and control) into complex systems. However, these frameworks primarily focus on system-level integration rather than controller design, leaving the implementation-specific challenges largely unaddressed.
Tools such as MATLAB/Simulink used by control engineers support controller development
and testing, but do not fully address the transition from design to software implementation. It does not help that control engineers often lack comprehensive knowledge of software quality assurance methods, and
software engineers have little understanding of control theory, be it controller design or discretization challenges. The lack of implementation guides exacerbates  the problem.   

Despite the critical role of controllers in robotics systems, research often overlooks their practical implementation. Studies focus on theoretical concepts, such as new controller architectures providing different guarantees for a specific system\,\cite{pan2025survey}, neglecting real-world controller deployments. Empirical studies of actual controller implementations are needed to bridge this gap, by investigating practical methodologies for robotic controller implementations and their quality assurance.
\looseness=-1
We present one such empirical study below, focusing on the ROS software ecosystem.
Specifically, we ask:\looseness=-1


\parhead{RQ1} \emph{What are the most common applications of
  controllers in robotic systems?}
We study robotic applications to obtain an overview of typical environments and tasks that controllers are confronted with and how these are solved from a control-theory perspective.
We investigate common controllers and their design.


\looseness=-1 \parhead{RQ2} \emph{How are controllers implemented and
  what are the characteristics of these implementations?}
We study the source-code of controller
implementations to infer how they handle
fundamentally different domains, from continuous- to discrete-time.
Furthermore, we analyze code encapsulation,
real-time handling (e.g., loop time), and safety-relevant
aspects, such as handling of outdated signal values.


\looseness -1
\parhead{RQ3} \emph{Which verification and validation techniques are used in controller implementations to ensure safety and reliability?}
The use of tools, such as MATLAB/Simulink, facilitates the evaluation of performance and reliability of controllers in a simulation environment. However, the implemented controller can differ in performance form the MATLAB/Simulink model, and requires additional verification and validation. We examine the controller implementations to identify common software quality practices, such as testing methods, that ensure safety and reliability.


\looseness=-1
We present an empirical study of 186 controller implementations in robotics systems from 141 GitHub repositories based on ROS, the largest robot software ecosystem and the most prominent robot middleware\,\cite{10.1145/3368089.3409743}.
Our findings indicate that the implemented controllers  implicitly assume con\-ti\-nuous-time behavior, handling discretization implicitly by delegating it to the middleware, often without any safety mechanisms, raising concerns about real-time reliability. The software test practices used are mostly superficial; we see no systematic verification of theoretical guarantees, leading to potential inconsistencies between expected and actual system behavior. These insights highlight the need for better implementation guidelines for controllers and rigorous verification techniques to enhance the reliability and safety of robot software controllers. Our dataset is available online.\footnote{\url{https://figshare.com/s/c886163d68174ef62a90}}

\section{Background}
\noindent
\looseness=-1
Control theory provides the theoretical foundation for controller design. It ensures correctness-by-design and provides guarantees for properties, such as stability, settling time, and steady-state error\,\cite{ogata2010modern}. It also accounts for robustness against external disturbances and system uncertainties.
Control engineers distinguish between feedback (closed-loop) and feed-forward (open-loop) control. In feedback control loops, the system's output is measured and fed back to the controller. This theoretical framework is crucial for developing self-adaptive systems that require reliable and robust control mechanisms\,\cite{weyns2020introduction}.
In contrast to feedback control, feed forward control does not use any self-adaptation mechanism and only considers a forward pass.
Designing a controller involves modelling the systems dynamic behavior and defining a control law. 

Control laws---blueprints of combinations of different algorithms---range from simple designs for single-input-single-output system (SISO), to complex ones for multiple-input-multiple-output systems (MIMO).
For example, model predictive control (MPC) relies on a model of the system (also called the plant) and formulates a constrained optimization problem solved at runtime. It is particularly well suited for controlling MIMO systems.
In contrast, proportional-integral-derivative (PID) controllers do not rely on a system model and employ a simple proportional, integral, and derivative structure, making them suitable for SISO systems.
Numerous control laws have been developed for specific applications, including PID, MPC, linear-quadratic regulators (LQR), and linear-quadratic-Gaussian regulators (LQG)\,\cite{GARCIA1989335, 8792042, 704994}.

Traditionally the system and the controller are both assumed to operate in the continuous domain. However, modern controllers are implemented in software\,\cite{ogata2010modern}, and, consequently, they perform discretization of the continuous variables.
While a broad theory of discrete controller design exist\,\cite{ogata1995discrete}, its application and implementation in real robotic systems remains unclear.
Many discretization approaches, such as Z-transformation\,\cite{Wiley}, rely on a fixed sampling time and assume constant loop execution.
Deviations from this assumption, for example due to variable execution times or system load, can lead to unexpected behavior, since the controller's assumption that control signals are applied at fixed intervals is violated.
If the system is subject to excessive load, it can exhibit jitter, making the maintenance of a constant frequency control loop difficult in practice.
Studies have shown that implicit or inappropriate discretization may violate guarantees provided by controller theory\,\cite{di2019sampled},\cite{cervin2003does}.
Different discretization methods have been studied in the literature; however, their underlying theory is often complex and difficult to implement.
Consequently, ensuring consistent timing and explicitly addressing discretization is an important aspect of controller implementation.

\begin{figure}[t]
	\centering
	\includegraphics[width=\columnwidth]{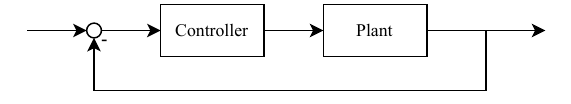}
\vspace{-0.7cm}
	\caption{Visualization of a control loop as a block diagram}
	\label{fig:ctLoop}
	\vspace{-0.5cm}
\end{figure}

\looseness=-1
To facilitate the comprehension of control laws and to provide a visual representation of the architecture, control engineers employ models of the control law in the form of block diagrams\,\cite{ogata2010modern}. Each block in a block diagram corresponds to a specific algorithm or operation, abstracting the flow of signals and calculations. \Cref{fig:ctLoop} presents a block diagram of a standard feedback loop.
MATLAB/Simulink uses block representations to model controllers, making it easy for control engineers to adapt due to the similarity. MATLAB/Simulink also offers tools such as Control System
and the Model Predictive Control toolbox,\footnote{\href{https://de.mathworks.com/products.html}{https://de.mathworks.com/products.html}} which simplify controller development with prebuilt blocks and tuning facilities.
In addition, these tools provide features to automatically generate
code from the block diagrams, whereby the discretization is implemented automatically.

Middleware platforms like ROS\,\cite{quigley2009ros,doi:10.1126/scirobotics.abm6074}, and OROCOS\,\cite{1242011} are widely used for developing robotic systems. A robotic software middleware is a framework that integrates the many software parts needed in a robot, enabling modular construction, largely (but not only) through configuration.
The middleware platforms provide libraries, such as ros\_control\,\cite{ros_control}, which focus on the integration of controllers rather than on controller design. The ROS2 project has published guidelines for writing custom controllers,\footnote{\href{https://control.ros.org/jazzy/doc/ros2_controllers/doc/writing_new_controller.html}{https://control.ros.org/jazzy/doc/ros2\_controllers/doc.html}} again, focusing on the integration aspects rather than on
the design.

\looseness=-1
The separation of expertise between control and software engineers hinders the development of control software, as each often overlooks the other's challenges. The fundamental differences between both fields further complicate cross-domain integration of quality practices\,\cite{7194659}. Although tools to automatically generate controller code exist, the evolution, maintenance, and especially integration of controller implementations into larger systems remains manual and error-prone\,\cite{Li2024}. Thus, understanding the controller implementations and quality practices around them is crucial for the engineering of safe and reliable robotic systems.


\section{Methodology}

We identified and analyzed controllers as follows.


\subsection{Controller Identification}\label{sec:methododology:selection}

We mined relevant open-source repositories of ROS-based robotics applications.  We focused on GitHub, home to the vast majority of open-source robotics projects. We excluded smaller platforms, such as GitLab and Bitbucket. For subject selection, we applied the following inclusion and exclusion criteria derived from our research questions and the specifics of the robotics domain.

\parhead{Inclusion Criteria.}
The following three criteria needed to hold to include a repository:

\emph{\emph{(I1)} The system uses ROS.}
We restricted our search to ROS, because it is closely related to the area we are interested in. We do not differentiate between ROS1 or ROS2, even though ROS1's end-of-life date was May 2025. 
Furthermore, we focused on projects that use general-purpose languages, such as C++ or Python for the controller implementation. Consequently, we did not consider MATLAB/Simulink, Arduino, or similar tools.

\emph{\emph{(I2)} The controlled system is in the robotics domain.}
We define a robot as an automated, self-adaptive machine capable of executing specific tasks with minimal human intervention and moving freely within a predefined space. Although ROS aligns closely with the robotics domain, this focus remains exclusively on robotics as previously defined. Projects involving controller implementations for e.g., wind turbines, are not relevant for the scope of this study.

\looseness=-1
\textit{(I3) the repository contains a controller implementation, uses control libraries, or packages utilizing controllers.}
We include full robotic systems and repositories containing a single controller. Repositories without controller implementations but using libraries or packages with controllers are also considered. In such cases, we analyze the controller integration and implementation of the library/package. The controller must control a robot according to \textit{(I2)}.

Based on the inclusion criteria, we defined two keywords as search terms: \textit{ROS} and \textit{Controller}. \textit{ROS} fulfills \textit{(I1)} and \textit{(I2)}, limiting the search to robotics-related repositories, while \textit{Controller} filters for controller-related repositories.
Since ROS1 is EOL in March 2025 we split our selection process for ROS1 and ROS2 and performed two searches:  Resulting in a search with the keywords \textit{ROS} + \textit{Controller} and \textit{ROS2} + \textit{Controller}.
For the dataset including ROS1, the number of stars was used as an additional inclusion criteria. We exclusively considered ROS1 repositories with at least 5 stars, as they are older and have the potential to achieve a higher number of stars.
After applying the keyword and stars criteria, GitHub suggested 450 repositories (ROS and ROS2 combined) for the time the study was conducted.
%
\looseness=-1
However, within the 450 repositories still numerous are irrelevant, therefore, we defined additional exclusion criteria to ensure the relevance.

\parhead{Exclusion Criteria.}
We excluded repositories when at least one of the following criteria applied:
\looseness=-1
\textit{(Ex1) The project is a driver.} 
Projects with drivers often only contain interfaces to specific hardware, not a controller implementation, nor make use of them. Therefore, do not help to answer the research questions.

\looseness=-1
\textit{(Ex2) The robot is controlled with a joystick or similarly.} 
We exclude these project because the actual control task is done by a human and not a controller.
%

\looseness=-1
\textit{(Ex3) No documentation.}
We did not expect large and extensive documentation, since we do not focus on industrial controller implementations only, which would be a study on its own. However, we expect a bear minimum, e.g., a small project description.

We applied the exclusion criteria to the 450 mined repositories, eliminating 309 and keeping 141 (78 ROS1-based, 63 ROS2-based) for the analysis.
The filtering of the repositories was primarily through a manual process.
Consequently, we manually reviewed the inclusion and exclusion criteria for each repository. In borderline cases, we discussed to reach a consensus. Additionally, we cross-checked samples of repositories to ensure consistency in applying the in- and exclusion criteria. The inclusion rate for the ROS and ROS2 datasets was 33~\% and 30~\%, respectively, indicating a comparable quality of repositories in the two datasets.

\subsection{Controller Analysis}
We analyzed our subject projects as follows. Specifically, we identified the most common applications for controllers and control laws in ROS (RQ1, \cref{sec:apps}) by investigating the controlled system and documentation.
Subsequently, to determine what control laws are used and how, we identified the controller code in the repositories and reviewed it to study controller design and implementation (RQ2, \cref{sec:impl}).
Finally, we investigated the verification and validation (V\&V) techniques (RQ3, \cref{sec:findings:rq3}) used to ensure the correct functionality of controller implementations.

\looseness=-1
All repositories were analyzed manually and qualitatively, primarily by the first author, who systematically inspected the code and reviewed documentation. He was an expert in control theory, supported by the co-authors, who contributed decades of software-engineering research expertise.
In case of doubt, uncertainties, or borderline cases, the first author consulted with the co-authors to resolve the issues through joint discussion and reach a consensus.

An important part was to identify the actual controller implementations in each codebase.
We searched in the codebase using keywords, typical directory structures, file naming conventions, C++ header files (to identify relevant methods), and dependent libraries and packages (for instance the Python control system library).\footnote{\href{https://github.com/python-control/python-control.git}{https://github.com/python-control/python-control.git}}
We additionally used ROS guidelines for custom controller development as a reference points to support consistent identification across projects.\footnote{\href{https://control.ros.org/jazzy/doc/ros2_controllers/doc/writing_new_controller.html}{https://control.ros.org/jazzy/doc/ros2\_controllers/doc/writing\_new\_controller}}
We searched for characteristic control loop and update methods (e.g., \texttt{update()}, \texttt{loop()}, \texttt{move()}, \texttt{step(),} \texttt{calcCmd()}), to qualitatively analyze the code. If a repository had multiple controllers, we considered each of them individually.
We identified 184 controller implementations in these repositories, divided into 102 ROS1 and 82 ROS2 controllers.

We answered our research questions as follows. For RQ1 (applications and control laws), we investigated the repositories regarding what systems are controlled to obtain a distribution of common applications (RQ1.1) and what control laws are commonly used (RQ1.2). This involved classifying controllers based on their application domain (e.g., mobile robots, manipulators, aerial systems) and extracting the control law by inspecting the code. For RQ2 (encapsulation and implementation), we studied the controller design and implementation.
For encapsulation (RQ2.1), we traced back the controller callback through the codebase, identifying when the controller code is called. Thereafter, we examined the code around the controller callback regarding timing and safety.
For controller implementation (RQ2.2), we analyzed the code in details.

%
For RQ3 (verification and validation), to understand how controller implementations ensure safety and reliability, we anticipated the differences in practices of robotics and regular software engineering, and took a broad view of testing, including both simulation-based testing (RQ3.1) and code-based testing (RQ3.2). Recall that robotics is a field dominated by mechanical and electrical engineers rather than software engineers\,\cite{10.1145/3195836.3195853}. While simulation-based testing is known to be common among mechanical and electrical engineers, code-based testing is an established practice in software engineering.
To identify V\&V techniques we examined the documentation and codebase for simulator and test artifacts.
\begin{table}[t]
	\centering
  \newcommand \tskip {1.5mm}
  \newcommand \thdr [1] {\textbf{\textsf{#1}}}
  \caption{The controlled systems\label{Tab.plants}}%
\vspace{-0.3cm}
  \begin{tabular}{>{\small}l>{\small}r}
  \textsf{category}      &  {\textsf{number of systems}} \\
	  \toprule
		\thdr{Mobile Robot} & \thdr{70}                                                                \\
		~~Two-wheeled                                 & 29\\
		~~Four-wheeled                                & 29                                                                                             \\
    ~~Omnidirectional                             & 12                                                                                              \\[\tskip]
	    \thdr{UAV}           & \thdr{33}                                                                \\
	    ~~Single                                      & 29                                                                                              \\
	    ~~Swarm                                       & 4                                                                                              \\[\tskip]
		\thdr{Manipulator}  & \thdr{28}                                                                \\
    \thdr {Humanoid} & \thdr 3 \\
    \thdr {Quadruped (Dog)} & \thdr 2 \\
    \thdr {Underwater} & \thdr 2 \\
		\bottomrule
	\end{tabular}
\vspace{-.1cm}
\end{table}

\section{Applications \& Control Laws (RQ1)}\label{sec:apps}
We now describe the detailed methodology and present the results for RQ1.1 (applications) and RQ2.2 (control laws).



\subsection{Detailed Methodology}\label{sec:methododology:apps}
\looseness=-1
To characterize the controlled system (the \emph{plant}) and application (RQ1.1), we checked whether the documentation found in the repository describes the controlled system or a use case. Then, we investigated the implementation artifacts: URDF and yaml files, and ROS odometry message definitions to identify the robot type.  URDF is a language for specifying robot's geometry, e.g., number of joints/wheels, wheel radii, and wheel separation. A lack of a URDF file indicates that the geometry/dynamic may be written in yaml files or published with odometry messages. Finally, we studied the controller code, as it can contain information about the controlled system. For example, if the implemented controller is a differential-drive controller, the controlled system is likely a two-wheeled robot.  After identifying the robots, we performed clustering according to the operational environment of the application.


\looseness=-1
For control law identification (RQ1.2), we examined the documentation searching for keywords PID, MPC, LQR, which correspond to specific control laws. If the documentation contained information about the domain, we searched for user-defined keywords that relate to the specific domain. The controller implementation guidelines we already used to identify controller code, were also instrumental here.
In case the documentation or keywords did not provide sufficient information, the control law was reversed engineered then mapped to existing ones, if possible. 
\subsection{Applications}

The analyzed systems are common for the robotics area and fall into three major categories: mobile robots, unmanned aerial vehicles (UAV), and manipulators (\cref{Tab.plants}).  We define mobile robots as wheeled bases without additional actuators and divide them further into two-wheeled with differential drive, four-wheeled with Ackermann steering, and omnidirectional with mecanum wheels.  We define UAVs as systems capable of controlled flight, regardless of the propulsion method (rotors or turbine), and divide the projects into concerning a single robot and swarms (multiple robots). Most UAVs controlled in GitHub projects are quadcopters. We classify stationary robotic arms as manipulators. For about half of the subject projects a particular model and make are identifiable, including Franka Emika Panda, Yasakawa, UR5e (and other UR variants), Staubli, and SCARA.  The small set of remaining projects (only seven in total) contains humanoid, quadruped and underwater robots (AUV). These are much less common, likely because they are much more complex to control (humanoid, quadruped) or significantly harder to deploy (underwater). Especially the humanoid and quadruped robots are relevant from the perspective of this study, as they do not follow standard control designs presented in classic textbooks.
%
%
%
%
%
%
We failed to identify the controlled system only in six of the studied repositories, of which most repositories use a PID controller and did not provide sufficient information for identification.

\begin{leftbar}
\noindent Most controlled systems belong in well-studied plants classes in control theory, allowing to follow an established controller design. Exceptions are humanoid and quadruped robots.
\end{leftbar}

\begin{table}[t]
	\centering
	\newcommand \tskip {1.5mm}
	\newcommand \thdr [1] {\textbf{\textsf{#1}}}
	\caption{The control laws\label{Tab.controller}}%
	\vspace{-0.3cm}
	\begin{tabular}{>{\small}l>{\small}r}

    \textsf{control law}   & \llap{\textsf{number of controllers}} \\
	  \toprule
		\thdr{Feedback loop control} & \textbf{155}                                                                \\
		~~PID                                         & 55                                                                                             \\
		~~Model-Predictive-Control                    & 21                                                                                              \\
		~~Cartesian Controller                        & 18                                                                                              \\
		~~Pure Pursuit                                & 12                                                                                              \\
		~~Linear-Quadratic-Regulator (LQR)            & 5                                                                                              \\
		~~Stanley                                     & 3                                                                                              \\
		~~Geometric Controller						& 3																								\\
		~~Linear-Quadratic-Gaussian-Regulator (LQG)   & 1                                                                                              \\
		~~Reinforcement Learning					    & 1                                                                                              \\
		~~Model-Reference-Adaptive-Control			& 1 																							\\
		~~Custom                                      & 35                                                                                             \\[\tskip]
		\thdr{Feed forward control}  & \textbf{29}                                                                \\
		~~Differential Drive                          & 9                                                                                             \\
		~~Ackermann Steering                          & 4                                                                                              \\
		~~Custom                                      & 16 \\
		\bottomrule
	\end{tabular}
\vspace{-.2cm}
\end{table}

\subsection{Control Laws}
The most frequently used feedback loop controller is PID (\cref{Tab.controller}) with 30~\%. This reflects its popularity in the industry, due to the simplicity of its design\,\cite{wescott2000pid}, while providing good stability and robustness properties\,\cite{1220756,GARPINGER2014568,wescott2000pid}.  It is followed by model-predictive controllers, Cartesian controllers (for manipulators), and Pure Pursuit.
Pure Pursuit, Stanley and Geometric are path tracking controllers using a simple kinematics model of the plant and state-feedback. Other model-based controllers, such as MPC, use a discrete model of the controlled system to predict the future behavior over a fixed prediction horizon\,\cite{schwenzer2021review}.
Feed forward controllers (16\%) map inputs, such as velocity or steering wheel angle, to the designed geometry and motor control signals.   The most common feed forward controllers follow two common kinematics: differential drive and Ackermann steering. The remaining 16 controllers have individual kinematics, such as for omnidirectional robots with various wheel configurations or mecanum wheels.
Only 35 out of 155 feedback loop controllers and 16 out of 29 feed forward controllers did not follow standard control laws and are marked as ``custom'' in \cref{Tab.controller}.

Most of the identified controllers (\cref{Tab.controller}) follow well researched designs. Nearly all address robot motion from simple position control to path following, whether through feed forward control or more advanced designs with fine-grained guarantees of disturbance rejection and robustness.
The repositories \ghrepo{Cartesian Controller}{https://github.com/fzi-forschungszentrum-informatik/cartesian_controllers} and \ghrepo{tswr\_awsim}{https://github.com/mgmike1011/tswr_awsim} contain examples of more complex controllers.
Controller labeled as \textit{custom} solve inverse kinematics problems combined with state feedback control, which is a common approach for manipulator control\,\cite{6413422}.
For AUV, controllers address attitude, position, or path-tracking control.
The distribution of control laws that we observe is consistent with the existing research\,\cite{ALBONICO2023111574}.
\looseness -1

\begin{leftbar}
	\noindent Most controlled systems follow standard control laws, however not always the most robust ones. The majority of identified control laws are designed for continuous systems.
\end{leftbar}

\noindent

\section{Implementations Characteristics (RQ2)}\label{sec:impl}
\noindent\looseness=-1
We now describe the detailed methodology and then the results on structure and encapsulating (RQ2.1), and implementation (RQ2.2).

\subsection{Detailed Methodology}\label{sec:methododology:impl}
\looseness=-1
We examined how controllers are implemented structurally and how they are encapsulated (RQ2.1), and how the control laws are actually implemented (RQ2.2). We focused on how and where the controller is executed from the system, as well as timing and safety aspects. We examined header and source files of the controller code, 
to determine the code structure of the controller implementation. 
Thereby, we determine whether the implementation is scattered over multiple methods or classes, or whether implemented modularized.
Furthermore, we investigate how the controller is implemented/encapsulated in ROS, regarding node and code structure.
ROS2 provides guidelines for writing custom controllers, including a standardized method structure and control loop call.
For ROS1 these guidelines do not exist.
We investigated to what extent these guidelines were followed.
Moreover, we identified the common programming languages used for controller implementations, by analyzing what are the most occurring file types, for instance, .cpp or .py. 

\looseness=-1
To analyze the controller implementation (RQ2.2), we examined how control laws are realized, specifically whether the equation is directly mapped from continuous domain or whether discretization methods are used.
One focus is on how controllers handle discretization for time and values, and how they implement timing and value constraints. We also analyzed controller trigger events and their frequency (e.g., whether fixed interval or event/condition-based), the real-time handling (e.g., loop time), and other safety-relevant aspects, such as outdated signal values. To study these aspects of the questions, we reviewed the controller code manually. Specifically, the first author, who has a background in control theory, studied the code and consulted with the other authors (experts in software engineering and robotics systems). To analyze the timing aspects, we especially looked for specific libraries used in the code, such as chrono for C++, as well as ROS's client libraries for C++ (rclcpp) and Python (rclpy), searching for relevant keywords, such as \texttt{rclcpp::Time} or \texttt{$^*$::Duration}.

\subsection{Implementations Structure \& Encapsulation}
\noindent
Our analysis revealed the following structures of the controller implementations (RQ2.1) and the following insights into the actual implementation of control laws in code (RQ2.2).

Recall that ROS2 provides guidelines for writing controllers (e.g., method structure). However, we found that only 19 projects out of 63 adhere closely to these guidelines.
Consequently, 44 ROS2 repositories do not follow the guideline with additional 78 ROS1 repositories that do not have any guidelines resulting in unique controller structure. While non-adherence or non-existing guidelines may not cause system failure, it can lead to arbitrary encapsulation.
When layers or wrappers are introduced without explicit responsibility, timing contracts, and lifecycle concerns (ROS nodes, parameters, executors) become entangled with numerical control logic and parameter handling is scattered across files. This complicates traceability and maintainability, making controllers harder to understand, and test. 
Furthermore, custom extra wrapper layers or ROS callbacks that only pass data may add executor and serialization overhead, which increases latency and jitter in control loops and complicates timing diagnostics.


We observed a strong alignment between the structure of the controller block diagrams and the implementation of the controllers.
It is noteworthy that many identified control laws, such as PID and LQR, are represented as single blocks in block diagrams. Maintaining this structure in the source code, without subdividing or consolidating into a single file, facilitates tracing each controller and its function. A notable example is the \ghrepo{Cartesian Controller}{https://github.com/fzi-forschungszentrum-informatik/cartesian_controllers} repository, where the source code structure, including files like ForwardDynamicsSolver.cpp, IKSolver.cpp, and PDController.cpp, mirrors the block diagram in the documentation. Similarly, the \ghrepo{lqgController}{https://github.com/HaoguangYang/lqgController} separates the LQR controller and Kalman filter into LQR.cpp and Kalman.cpp, respectively, reflecting the block diagram.
For repositories using MPC this also holds with rear exceptions. MPC usually consists of two parts, a model of the control system and an optimizer solving the optimization problem \cite{schwenzer2021review}. For the solver external libraries are used, such as Ipopt.
While this correspondence between block diagrams and source code can improve interpretability for control engineers, it has inherent limitations. Encapsulation in ROS determines when and how controller code is executed, including its interaction with callbacks, executors, parameters, and lifecycle states. If these execution boundaries are poorly defined, a correct controller implementation remains ineffective regardless of how closely the code mirrors a block diagram. Moreover, block-diagram representations are often unfamiliar to software engineers, limiting their usefulness as a general traceability mechanism. Consequently, clear encapsulation is a prerequisite for traceability and correctness, whereas block-diagram alignment should be regarded as a secondary, supportive practice.

We identified two methods for implementing and utilizing controller code in our dataset.
The first method involves encapsulating the controller within a single file, which includes necessary methods such as calculating control signals and actuator limitations (e.g., limited acceleration). The controller is called from another file that gathers current sensor data, passing this data to the controller method as arguments, which then calculates and returns control signals for the actuators. An example is the \ghrepo{flock2}{https://github.com/clydemcqueen/flock2} repository, where a PID controller is implemented in pid.hpp, and the method call occurs in flight\_controller\_simple.cpp within the \texttt{\_odom\_callback()}. method.
The second method involves conducting all operations within a single file. Here, the controller constitutes a single file or class, often named according to its purpose. For instance, in the \ghrepo{Cartesian Controller}{https://github.com/fzi-forschungszentrum-informatik/cartesian_controllers} repository for manipulators, the calculation is implemented in the \texttt{computeComplianceError()} method in cartesia\_compliance\_controller.cpp, and this method is called within the \texttt{update()} method in the same file.

\begin{leftbar}
\noindent\looseness=-1 Arbitrary encapsulation of controller code is a primary obstacle to traceability and maintainability in ROS-based controllers. Poorly defined wrapper layers entangle control logic with execution, lifecycle, and timing concerns, undermining correct controller execution even when the control law itself is well implemented.
\end{leftbar}

\subsection{Controller Implementations}
We identified several implementation strategies related to time handling, discretization, and derivative computation.

Time handling varied considerably depending on the programming language and underlying control law. In model-free controllers, as specially PID, we observed the use of internal clocks implemented as local timers within a control loop. These timers measure the interval between two control-loop iterations and are used to compute time-dependent terms. An example of an internal clock is contained in the repository \ghrepo{UTNuclearRobotics}{https://github.com/UTNuclearRoboticsPublic/pid} contains a PID controller implementation. Listing~\ref{lst:pidTime} shows a snippet of the source code with the implemented internal clock.
%

\begin{lstlisting}[caption={Time handling of the PID controller (internal clock)}, float=b, label=lst:pidTime, mathescape, belowskip=-0.0cm, aboveskip=-0.1cm]
...
// calculate delta_t
	delta_t_ = this->now() - prev_time_;
	prev_time_ = this->now();
...
\end{lstlisting}

Explicit discretization methods where absent in analyzed model-free controller implementations.
Only numerical approximations of the derivative were found, using finite differences. 
Despite the known sensitivity of derivative computation to jitter and sampling irregularities, we did not observe dedicated mechanisms to mitigate these effects, such as derivative filtering or adaptive time-step handling.
For integration, we also did not find any dedicated methods, however, integration calculation is less sensitive to jitter and primarily change the integration window only.
%
Consequently, safeguards are needed to ensure proper controller behavior.
However, we did not observe safeguards addressing discretization errors or timing inconsistencies, only basic value checks.
These typically consisted of conditional statements that prevent invalid values (e.g., NaN, or infinite), as illustrated in Listing\,\ref{lst:invalidValues}. 

\begin{lstlisting}[caption={Safeguards for invalid values}, float=t, label=lst:invalidValues, mathescape, belowskip=-.6cm,aboveskip=+.3cm]
...
if (dt == 0 || $\textcolor{blue}{\textbf{std::isnan}}$(error) || $\textcolor{blue}{\textbf{std::isinf}}$(error) || $\textcolor{blue}{\textbf{std::isnan}}$(error_dot) || $\textcolor{blue}{\textbf{std::isinf}}$(error_dot)) {
	return 0.0;
} ...
\end{lstlisting}

Controller-inherent techniques, such as anti-windup, were not considered. Anti-windup mechanisms are commonly used in practice to prevent integrator saturation under actuator limits, we excluded hardware-dependent constraints from our analysis and focused solely on software-level implementations.
Most PID controller implementations exposed interfaces for gain adjustments (e.g., proportional, integral, derivative), typically through getter and setter methods, enabling runtime tuning.
\looseness=-1


We observed clear differences in the implementation strategies when comparing model-based controllers, such as MPC and feed forward controllers, to model-free controllers.
These controllers rely on system models and internal prediction mechanisms, which strongly influence their implementation.
%
For MPC implementations, the internal simulations were executed using a fixed time step that is decoupled from the external control-loop frequency.
As a result, the real-time capabilities depend on the ability of the optimization and simulation component to complete within a specific time window. 
Across the analyzed repositories, explicit enforcement of solver time constraints was rare. We identified only a single implementation that explicitly specified an upper bound on solver runtime \ghrepo{Hypharos\_Minicar}{https://github.com/Hypha-ROS/hypharos_minicar/blob/master/src/MPC.cpp} line 312.
In all other repositories, real-time capabilities was implicitly assumed rather than explicitly verified.
Prediction models used within MPC implementations were discretized according to the fixed internal simulation step size.
This discretization was generally independent of the ROS control-loop scheduling, indicating a separation between prediction model timing and controller execution timing.

We observed analogous strategies in feed forward controllers, such as differential drive and Ackermann steering controllers.
These controllers receive linear and/or angular velocity inputs and decompose them into wheel velocities.
Odometry is updated either from motor feedback or velocity commands.
To ensure dynamic integrity and odometry precision, the controller incorporates a timeout safeguard, causing the robot to cease operation if breached. A snippet illustrating this strategy is shown in Listing\,\ref{lst:timeConstraints}.
Model-based controllers in a feedback loop offer limited possibilities to adjust controller parameters and the system model. While minor parameter changes are feasible, substantial modifications, such as altering system dynamics or prediction models, require source code changes. Feed forward controllers often allow changes to the robot's geometry, such as wheel radius or wheel separation.

\begin{lstlisting}[caption={Safeguard for values within time window}, float=b, label=lst:timeConstraints, belowskip=-0.0cm, aboveskip=-0.05cm]
...
const auto age_of_last_command = time - last_command_msg->header.stamp;
// Brake if cmd_vel has timeout, ...
if (age_of_last_command > cmd_vel_timeout_) {
	last_command_msg->twist.linear.x = 0.0;
	last_command_msg->twist.linear.y = 0.0;
	last_command_msg->twist.angular.z = 0.0;
} ...
\end{lstlisting}

\begin{leftbar}
\noindent\looseness=-1 The findings of RQ2 show that the implementation of critical elements appears to be \textit{ad hoc}. Implementations rely on basic numerical approximations and internal timing mechanisms, while explicit discretization methods are not applied.
Safeguards address upper limits for control-loop times or ensure that actuator signals are not outdated. Value safeguards ensure that calculated errors or control-loop times are real numbers. Therefore, \textbf{real-time capability and proper timed control signals remain questionable}, which raises whether the guarantees from control theory, still hold.
\end{leftbar}


\section{Verification and Validation (RQ3)}\label{sec:findings:rq3}
\noindent
Following the analysis of controller implementation practices in RQ2, we investigated the verification and validation techniques used to assess controller correctness and reliability.
\looseness=-1
\subsection{Detailed Methodology}%
\label{sec:methododology:vv}

\noindent\looseness=-1
We categorized testing techniques into three groups: simulation-based testing, code-based testing, and unknown.
\@ \emph{Simulation-based testing} (RQ3.1) refers to the evaluation of robot behavior and control algorithms in virtual environments that approximate real-world conditions.
Simulation-based testing is often ad hoc and manually executed, therefore, we optimistically interpreted the presence of simulator configurations or launch files in a repository as evidence of at least ad hoc simulation-based testing.
\@ \emph{Code-based testing} (RQ3.2) refers to automated testing approaches applied directly to the codebase, including unit tests, integration tests, and test frameworks, to detect errors early in development.
While code-based testing is inherently automated, it may still be ad hoc or vary in rigor depending on code or requirements coverage.
\@ \emph{Unknown} was assigned when no explicit evidence of testing was found in the codebase.
Although real-world tests may have been performed, such activities cannot be assessed in an artifact-based study if they are not reflected in the software infrastructure.
We examined the directory structure and codebase, searching for test folder or test cases and manually investigated the documentation and codebase regarding documented testing and simulator plugins.

\looseness=-1
To identify simulation-based testing (RQ3.1), we searched documentation and source code for simulator integrations and interfaces, including Gazebo plugins, configuration files, as well as external artifacts such as Simulink models (slx). 
We further analyzed the documentation to assess whether simulation usage followed a structured evaluation process or consisted of informal testing runs, determining the scope of the tests performed.

\looseness=-1
%
To identify code-based testing techniques (RQ3.2), we searched for dedicated test directories, test files, and testing frameworks. This included keyword-based searches adapted to the programming language used (e.g., \texttt{assert(...)}, \texttt{ASSERT\_EQ}, or \texttt{Test(...)}). Particular attention was given to tests that directly assess controller behavior, timing constraints, and correctness of control outputs.

\begin{table}[t]
	\centering
	\newcommand \tskip {1.5mm}
	\newcommand \thdr [1] {\textbf{\textsf{#1}}}
	\caption{Distribution of the identified V\&V techniques\label{Tab.V&V-techniques}}%
	\vspace{-0.3cm}
  \begin{tabular}{>{\small}l>{\small}r}

		\textsf{verification \& validation technique}      & \textsf{number of systems} \\
		\toprule
		Simulation-based testing      & \textbf{78}                                  \\
		Unit testing (code-based)      & \textbf{18}                                   \\
		Integration testing (code-based) & \textbf{7}                                   \\
		\bottomrule
	\end{tabular}

  \vspace{-.2cm}

\end{table}

\subsection{Simulation-based Testing}

\looseness=-1
We found that 90 out of the 141 studied repositories (64~\%) use simulation-based or code-based testing. The remaining 51 repositories did not provide any evidence of testing and are therefore counted as \emph{unknown}. Simulation-based testing was conducted 78 times and code-based testing 25 times, divided into unit testing 18 times and integration testing 7 times, see \cref{Tab.V&V-techniques}. Some repositories used more than one testing technique, therefore, the total number of techniques found is higher than the found repositories.

Simulation-based testing was the most frequent observed verification technique.
The analyzed repositories made use of several robotics simulators, including Gazebo, RViz, MuJoCo, CoppeliaSim, SVL Simulator, and AWSIM.
These repositories provided configuration files or launch scripts to start a simulation environment.
However, systematic use of simulation for automated controller validation and controller behavior was missing.
The majority of projects relied on manual simulation runs supported by visualization tools, indicating that the evaluation was primarily conducted through visual inspiration of the robot behavior.
Only one repository provided predefined scenarios or scripted trajectories for repeated evaluation.
The repository \ghrepo{Trajectory Tracking}{https://github.com/yabdulra/Trajectory-tracking-with-obstacle-avoidance} provides three predefined test scenarios, circular, eight-shape, and linear, which can be found under \texttt{trajectory\_tracking/trajectories}.
However, explicit test oracles, and coverage criteria were absent. 
This results in presumable large untested parts of the controller code, questionable controller performance, and especially questionable controller properties that are given by the control law but are not ensured in the implemented code.
Even though the engineering community is fond of using simulation in their work, the tool itself does not appear to be used systematically to do the V\&V work in the studied repositories, but rather as an exploratory debugging and demonstration tool. This appears to be a missed opportunity.


\looseness=-1

\begin{leftbar}
  \noindent
	Simulation-based testing is used superficially and ad hoc. The repositories show no evidence of a systematic coverage of the operational space, requirements, or properties. Evaluation appears to be done by a visual inspection, and not confronted with the control-theoretical models, or defined test oracles.
  \looseness -1
\end{leftbar}


\subsection{Code-Based Testing}
\looseness=-1
Code-based testing was observed less frequently (25).
Unit tests primarily focused on basic functionality, such as parameter initialization, signal validity checks, and fail-safe behavior when outdated inputs were detected.
An example of a unit tests for  parameters initialization is shown in \cref{lst:unitTestEx}, retrieved from the repository \ghrepo{clearpathrobotics}{https://github.com/clearpathrobotics/clearpath_mecanum_drive_controller}.
Tests targeting the dynamic behavior of controllers or the correctness of control outputs over time were not found.
Integration tests were mainly used to verify controller localization and loading.
In ROS2 repositories these tests were more prevalent than in ROS1, which highlights the usefulness of standardized interfaces and guidelines for integration tests as provided by ROS2.
%
Across the analyzed repositories, we did not observe the use of software verification techniques such as model checking or runtime monitoring\,\cite{zheng2014state, mandrioli2023stress, krichen2023survey}. 
%
%
\begin{lstlisting}[caption={Unit test example}, float=t, label=lst:unitTestEx, mathescape, belowskip=-.5cm, aboveskip=+.3cm]
$\textcolor{violet}{\textbf{TEST\_F}}$(MecanumDriveControllerTest, when_controller_is_configured_expect_all_
	  parameters_set)		{
	SetUpController();
	$\textcolor{blue}{\textbf{ASSERT\_EQ}}$(controller_->params_.reference_timeout, 0.0);
	$\textcolor{blue}{\textbf{ASSERT\_EQ}}$(controller_->params_.kinematics.
	wheels_radius, 0.0);
	$\textcolor{blue}{\textbf{ASSERT\_EQ}}$(controller_->params_.kinematics.
		sum_of_robot_center_projection_on_X_Y_axis, 0.0);
	...
}
\end{lstlisting}
%
\begin{leftbar}
  \noindent
	Code-based testing is conducted superficially and rarely in our dataset (18~\% of repositories).  Tests rarely address the core controller behavior or the controller integration into the system.
\end{leftbar}

The observed verification and validation practices indicate a strong reliance on manual simulation-based evaluation and a limited adoption of systematic, automated testing techniques. This suggests that controller correctness is often assessed at late development stages and primarily through informal validation rather than through continuous, systematically verification processes.

\section{Discussion}

\noindent
\looseness=-1 
Our study highlights the gap between theoretical control design and practical implementation in robotics software. 
While most controller (>~70~\%) implement well-established control laws, their implementation lack systematic handling of timing, discretization, and verification, which are essential for ensuring theoretical guarantees in real-world deployments.

\subsection{Control Theory -- Implementation Gap}

\parhead{Design Process.}
%
\looseness=-1
The majority of controlled systems and controllers identified in our dataset fall into well-studied categories, such as two-wheeled\,\cite{malu2014kinematics, 937419}, four-wheeled\,\cite{higuchi1993optimal}, or omnidirectional robots\,\cite{SongOmni}, for which mature control design methodologies exist. Consequently, controller design often follows well-established textbook procedures. However, these textbook procedures assume idealized execution environments and continuous environments instead of discrete environments as in software.

In particular, we observed that continuous control laws are often mapped directly into code without explicit discretization strategies or sampling-time design. Instead, controllers frequently rely on implicit loop timing provided by the ROS execution environment or internal software timers.
Therefore, dedicated discretization methods, such as Z-transformation, were absent, although, prior work has shown that such mismatches between implementation and continuous control laws can significantly affect performance and does not preserver the same properties\,\cite{di2019sampled, cervin2003does}.
Predictive controllers inherently operate on discretized models, however, explicit consideration of solver execution time and real-time constraints was rarely documented or enforced.

\looseness=-1
Furthermore, hardware-related constraints, such as sensor update rates, and communication delays, are considered during controller selection but were not reflected explicitly in the implementation. While a detailed evaluation of real-time performance is beyond the scope of this study, our observation suggests that many implementations implicitly assume ideal execution conditions, creating a mismatch between theoretical controller design assumptions and deployed robotic systems.

%

\parhead{Required Engineering Expertise.} 
\looseness=-1 
%
Our findings indicate that most controllers rely on well-established and experimentally tunable control laws, in particular PID controllers. This supports the observation that such controllers can be designed and tuned with limited modeling effort and modest control-theoretic expertise in practice\,\cite{wescott2000pid}. However, preserving control-theoretic properties in software implementations requires additional engineering expertise that goes beyond controller design alone.
Specifically, reliable controller implementation requires understanding the interaction between control algorithms, software execution, timing behavior, and real-time system constraints. Our analysis indicates that these aspects are often handled implicitly or in an \emph{ad hoc} manner, despite their known influence on performance\,\cite{cervin2002feedback, buttazzo2024hard}.
From a software engineering perspective, controller implementations often exhibit arbitrary encapsulation and limited adherence to established design patterns.
Controller code was frequently integrated into ROS applications without consistent encapsulation patterns or clear architectural conventions,  complicating reuse, testing, and long-term maintenance.
Our results suggest that controller engineering in robotics is not only a control-theoretic problem, but also a software engineering problem, requiring coordinated consideration of numerical computation, real-time execution, and system integration.

\looseness=-1
\parhead{Verification and Validation.}
\looseness=-1
The analysis of verification and validation practices further illustrates this gap between theory and implementation. Most testing activities focused on basic functional aspects.
Although control theory often provides formal guarantees and ``correctness-by-design,'' our findings indicate that these guarantees are not verified in the software implementation.
Simulation-based testing was predominantly used as a manual exploration and debugging tool rather than as a structured verification process. Automated evaluation, systematic scenario coverage, and explicit test oracles derived from control theory, were absent. Instead, simulation was employed for visual inspection and ad hoc experimentation, which limits reproducibility and provides weak evidence of correctness under varying conditions.

Similarly, code-based testing remained at a rudimentary level. Unit tests focused on basic functionalities, such as parameter initialization, getter and setter methods, or fail-safe conditions, while tests targeting the dynamic behavior of controllers were not observed. Although validating the dynamic behavior of controllers at the code level is challenging, the absence of behavioral tests indicates that controller correctness is often assumed rather than systematically assessed.
While traditional software engineering metrics such as code coverage do not measure control performance, they can provide a baseline. However, none of the analyzed repositories provided coverage reports or evidence of systematic test measurements. The absence further reduces confidence in the robustness of controller implementations, especially when reused across projects.
Advanced verification techniques, such as model checking or runtime monitoring\,\cite{zheng2014state, mandrioli2023stress, krichen2023survey}, were not observed. This suggests that these techniques have not yet been widely adopted in robotics software development, either due to limited tool support, integration barriers, or the lack of accessible workflows that connect formal methods with practical robotics development environments.

The limited adoption of systematic verification and validation techniques indicates that current controller implementations often rely on informal validation practices, or tests conducted in later stages of the development, e.g., field tests. Verification and validation of the controllers in the field, complicate the root-cause analysis; is the controller code, the integration, or other parts wrong. As a result, the theoretical guarantees remain uncertain in deployed robotic systems. While the controller design is well understood, software engineering aspects, such as verification, validation, and real-time constraints, lack systematic guidelines.
Especially with the growth of AI-based controllers (e.g., reinforcement-learning-based control), guidelines, best practices, and especially explicitly stated properties are needed to verify the behavior.
\subsection{Implications}
\looseness=-1 
While synthesizing best practices for correctly implementing controllers is beyond the scope of our work, we suggest actions for industrial practitioners, tool builders, and
researchers to improve real-world controller engineering.


\parhead{Industrial Practitioners and Developers.}
For industrial practitioners and developers, our results highlight that controller implementations are often treated as black boxes whose correctness is implicitly assumed based on control-theoretic design. In practice, however, implementation issues such as timing jitter, discretization, and missing safeguards can significantly degrade system performance and stability.
Therefore, practitioners should treat controller implementations as safety-critical software components.
More emphasis should be placed on unit testing and simulation-based verification to ensure the controller's reliability and safety, especially in complex environments.
This includes explicitly specifying control-loop timing assumptions, validating behavior under timing disturbances, and systematically testing fail-safe mechanisms such as timeouts and invalid signal handling.
While in practice controllers are tailored to specific hardware, making testing challenging, completely neglecting software testing can lead to arbitrary and unforeseen behavior in real-world operation.
%


\parhead{Tool Builders.}
For tool builders, our findings highlight opportunities to improve developer support for controller implementation and validation. The majority of controllers identified in our dataset address well-studied tasks, such as position and motion control for wheeled and omnidirectional robots, for which extensive control-theoretic knowledge already exists. Therefore, tools could provide structured guidance by recommending suitable controllers based on the underlying system model, together with documented trade-offs and implementation constraints.
Further tooling could embed stronger implementation support by providing recommended discretization methods, control-loop scheduling templates, and controller skeletons with built-in timing instrumentation. Such features would help developers avoid common mistakes related to time handling and execution jitter.
In addition, verification support should be priority rather than an optional activity. Providing test templates that target controller properties (e.g., response time, stability), instead of focusing on controller integration as currently emphasized by ROS2 guidelines, would lower the barrier for systematic testing. As demonstrated in large-scale industrial practice, even lightweight testing guidance can substantially increase testing adoption and awareness\,\cite{SEatGoogle}.
Finally, integrated runtime monitoring of properties would enable early detection of real-time violations during development enabling simulation tests become more meaningful.

\parhead{Researchers.}
For researchers, our study shows that advances in control theory does not automatically translate into reliable implementations. That is, research should be conducted on controller
implementations and integration, that bridges this gap.
In particular, scalable techniques for testing closed-loop behavior, runtime monitoring of safety properties, and systematic handling of timing uncertainties need further research. Addressing these challenges would significantly improve the quality of controllers, especially, with the growing of AI-based controllers.

\section{Threats to Validity}
\looseness=-1
%
\parhead{Internal Validity.}
The reliability of our findings may be compromised by errors introduced during the analysis of the source code. 
Manual analyses are inherent more prone to errors than automated analyses.
However, since only a minority of implementations adhere to the ROS guideline, the majority are unique, which makes automation challenging and leads to errors in special cases.
Although there is a possibility that keywords in the source code or errors in the analysis may be overlooked, we have mitigated this risk by discussing and cross-checking the results among all authors, especially when there is uncertainty about implementation details or documentation. 
Our dataset comprises two merged datasets related to ROS and ROS2 with a very similar inclusion rate (33~\% and 30~\%), showing a similar quality of both datasets.

\looseness=-1
\parhead{External Validity.}
The list of identified projects likely lacks projects from GitLab or Bitbucket, both of which are utilized in the robotics community. However, as previously stated, neither platform offers a code search, which has made it challenging to locate projects aligned with our area of interest. 
However, empirical research guidelines emphasize that it is more important to have a well-defined and well-scoped population (here: GitHub repositories with ROS-based projects) than trying to cover too many data sources, especially when the latter cannot be done systematically\,\cite{Ralph21_empirical_standards}.
It should be noted that the notion of completeness or representativeness is not applicable to empirical research, as the entire population of subjects is not known.
Additionally, it should be acknowledged that other frameworks, such as OROCOS, have not been considered in this study. It is possible that they provide further insight or reveal other insights into the characteristics of controller implementations.
\looseness -1
%

\section{Related Work}
\noindent
\looseness=-1
Control theory provides a rich theory for controller design and analysis, with literature covering stability, robustness, and performance guarantees\,\cite{trentelman2002control, bubnicki2005modern, glad2018control, 6313163, rivera1986internal}. However, these works do not address how controllers are realized, and tested in robotics systems. Bridging this gap between control theory and software engineering practice has therefore become an active research topic.

\citet{ALBONICO2023111574} conducted a systematic mapping study of software engineering practices in ROS-based systems. While their work provides a broad overview of research trends and tool support, our study complements this perspective by performing an empirical analysis of controller implementations. In particular, our RQ1 aligns with their investigation of robot types (RQ1.2), and our findings confirm the prevalence of wheeled robots and manipulators while extending the analysis to implementation and verification practices.
High-level challenges in robotic system design, including controller development and integration, have been studied\,\cite{10.5555/3283535.3283545}. These works primarily address architectural and conceptual issues, whereas our study focuses on concrete implementation practices.

Several approaches aim to formalize control-theoretic properties using software engineering techniques, for example by expressing guarantees in temporal logic\,\cite{9599335, 10.1145/3387939.3391568}. Similarly, behavior-tree-based frameworks have been proposed for structuring robot control and mission logic\,\cite{10.1145/3426425.3426942}. While these approaches target higher-level control or formal specifications, our work investigates how low-level controllers are implemented and validated in practice.

Model-driven development has been widely explored as a means to improve controller engineering. Tools such as RoboChart and RoboTool support state-machine-based modeling, automated code generation, and formal verification\,\cite{Li2024}. However, these approaches focus on idealized workflows and generated artifacts. In contrast, our study examines manually implemented controllers integrated into ROS-based systems, providing insight into how controllers are engineered outside controlled toolchain environments.

Software engineering research has also investigated model-based controller development using MATLAB/Simulink. Prior work proposes workflow improvements and quality analysis techniques\,\cite{pantelic2018software, liu2017improving, deissenboeck2008clone}. Reproducibility studies report limited accessibility of tools and models, resulting in replication rates below 10\%\,\cite{boll2020replicability}. Curated Simulink corpora further reveal limited reuse and maintainability of models\,\cite{chowdhury2018curated, boll2021characteristics, shrestha2023evosl}. While these studies analyze model-level artifacts, our work shifts the focus to implementation-level controller code.

\section{Conclusion}
\looseness=-1

\noindent
We conducted an empirical study of controller implementations in robotic systems, providing an overview of common application and analyzing implementation characteristics related to discretization, safeguards, and verification and validation techniques. By analyzing controllers sourced from GitHub, our work offers insights into real-world engineering practices that go beyond theoretical perspectives from control theory and software engineering.
This study helps bridge the gap between control-theoretic design and deployed robotic systems by characterizing how controllers are implemented. Highlighting recurring implementation patterns and challenges provides actionable insights.
%
%

As future work, we plan to extend our dataset to additional platforms such as Bitbucket and GitLab and broaden the search strategy.
We also intend to complement our quantitative analysis with qualitative studies, including interviews with industry engineers. However, this requiring a different methodology. In particular, we aim to further investigate verification and validation practices and controller development tool chains to improve awareness and adoption of systematic testing. Such follow-up studies could enable the development of practical guidelines, including patterns for unit testing and safeguard verification, and provide industry-driven feedback to the ROS community and robotics software ecosystem.




\bibliographystyle{ACM-Reference-Format}

\bibliography{references}

@STRING{feb = "Feb."}

@STRING{jul = "July"}

@STRING{oct = "Oct."}

@STRING{dec = "Dec."}

@inproceedings{chowdhury2018curated,
	author = {Chowdhury, Shafiul Azam and Varghese, Lina Sera and Mohian, Soumik and Johnson, Taylor T. and Csallner, Christoph},
	title = {A curated corpus of simulink models for model-based empirical studies},
	year = {2018},
	isbn = {9781450357289},
	publisher = {Association for Computing Machinery},
	address = {New York, NY, USA},
	url = {https://doi.org/10.1145/3196478.3196484},
	doi = {10.1145/3196478.3196484},
	booktitle = {Proceedings of the 4th International Workshop on Software Engineering for Smart Cyber-Physical Systems},
	pages = {45–48},
	numpages = {4},
	location = {Gothenburg, Sweden},
	series = {SEsCPS '18}
}

@article{boll2021characteristics,
	title={Characteristics, potentials, and limitations of open-source Simulink projects for empirical research},
	author={Boll, Alexander and Brokhausen, Florian and Amorim, Tiago and Kehrer, Timo and Vogelsang, Andreas},
	journal={Software and Systems Modeling},
	volume={20},
	number={6},
	pages={2111--2130},
	year={2021},
	publisher={Springer}
}

@inproceedings{boll2020replicability,
	author="Boll, Alexander
	and Kehrer, Timo",
	editor="Babur, {\"O}nder
	and Denil, Joachim
	and Vogel-Heuser, Birgit",
	title="On the Replicability of Experimental Tool Evaluations in Model-Based Development",
	booktitle="Systems Modelling and Management",
	year="2020",
	publisher="Springer International Publishing",
	address="Cham",
	pages="111--130",
	isbn="978-3-030-58167-1"
}

@inproceedings{shrestha2023evosl,
	author={Shrestha, Sohil Lal and Boll, Alexander and Chowdhury, Shafiul Azam and Kehrer, Timo and Csallner, Christoph},
	booktitle={2023 ACM/IEEE 26th International Conference on Model Driven Engineering Languages and Systems (MODELS)}, 
	title={EvoSL: A Large Open-Source Corpus of Changes in Simulink Models \& Projects}, 
	year={2023},
	volume={},
	number={},
	pages={273-284},
	doi={10.1109/MODELS58315.2023.00024},
	ISSN={979-8-3503-2480-8},
	month={Oct},
	publisher={IEEE},
	address={Västerås, Sweden },
}

@inproceedings{deissenboeck2008clone,
	author = {Deissenboeck, Florian and Hummel, Benjamin and J\"{u}rgens, Elmar and Sch\"{a}tz, Bernhard and Wagner, Stefan and Girard, Jean-Fran\c{c}ois and Teuchert, Stefan},
	title = {Clone detection in automotive model-based development},
	year = {2008},
	isbn = {9781605580791},
	publisher = {Association for Computing Machinery},
	address = {New York, NY, USA},
	url = {https://doi.org/10.1145/1368088.1368172},
	doi = {10.1145/1368088.1368172},
	booktitle = {Proceedings of the 30th International Conference on Software Engineering},
	pages = {603–612},
	numpages = {10},
	location = {Leipzig, Germany},
	series = {ICSE '08}
}

@inproceedings{liu2017improving,
	author={Liu, Bing and Lucia and Nejati, Shiva and Briand, Lionel C.},
	booktitle={2017 IEEE 24th International Conference on Software Analysis, Evolution and Reengineering (SANER)}, 
	title={Improving fault localization for Simulink models using search-based testing and prediction models}, 
	year={2017},
	publisher={IEEE},
	volume={},
	number={},
	pages={359-370},
	doi={10.1109/SANER.2017.7884636},
	ISSN={978-1-5090-5501-2},
	month={Feb},
	address={Klagenfurt, Austria },
	}

@article{pantelic2018software,
	title={Software engineering practices and Simulink: bridging the gap},
	author={Pantelic, Vera and Postma, Steven and Lawford, Mark and Jaskolka, Monika and Mackenzie, Bennett and Korobkine, Alexandre and Bender, Marc and Ong, Jeff and Marks, Gordon and Wassyng, Alan},
	journal={International Journal on Software Tools for Technology Transfer},
	volume={20},
	pages={95--117},
	year={2018},
	publisher={Springer}
}

@Book{ogata2010modern,
  author = {Ogata, Katsuhiko},
  title  = {Modern control engineering},
  edition = 5,
  year   = {2010},
  publisher={Pearson},
  address={Upper Saddle River},
}

@article{ros_control,
	TITLE = {{ros\_control: A generic and simple control framework for ROS}},
	AUTHOR = {Chitta, Sachin and Marder-Eppstein, Eitan and Meeussen, Wim and Pradeep, Vijay and Rodr{\'i}guez Tsouroukdissian, Adolfo and Bohren, Jonathan and Coleman, David and Magyar, Bence and Raiola, Gennaro and L{\"u}dtke, Mathias and Fernandez Perdomo, Enrique},
	URL = {https://hal.science/hal-01662418},
	JOURNAL = {{Journal of Open Source Software}},
	PUBLISHER = {{Open Journals}},
	VOLUME = {2},
	NUMBER = {20},
	PAGES = {456 - 456},
	YEAR = {2017},
	MONTH = Dec,
	DOI = {10.21105/joss.00456},
	HAL_ID = {hal-01662418},
	HAL_VERSION = {v1},
}

@article{wescott2000pid,
	title={PID without a PhD},
	author={Wescott, Tim},
	journal={Embedded Systems Programming},
	volume={13},
	number={11},
	pages={1--7},
	year={2000}
}

@inproceedings{ROS1,
	author = {Quigley, Morgan and Conley, Ken and Gerkey, Brian and Faust, Josh and Foote, Tully and Leibs, Jeremy and Wheeler, Rob and Ng, Andrew},
	year = {2009},
	month = {01},
	pages = {5},
	booktitle={ICRA workshop on open source software},
	title = {ROS: an open-source Robot Operating System},
	volume = {3},
	journal = {ICRA Workshop on Open Source Software},
	organization={Kobe, Japan}
}

@inproceedings{ROS2,
	author = {Maruyama, Yuya and Kato, Shinpei and Azumi, Takuya},
	title = {Exploring the performance of ROS2},
	year = {2016},
	isbn = {9781450344852},
	publisher = {Association for Computing Machinery},
	address = {New York, NY, USA},
	url = {https://doi.org/10.1145/2968478.2968502},
	doi = {10.1145/2968478.2968502},
	booktitle = {Proceedings of the 13th International Conference on Embedded Software},
	articleno = {5},
	numpages = {10},
	location = {Pittsburgh, Pennsylvania},
	series = {EMSOFT '16}
}

@INPROCEEDINGS{1242011,
	author={Bruyninckx, Herman and Soetens, Peter and Koninckx, Bob},
	booktitle={2003 IEEE International Conference on Robotics and Automation (Cat. No.03CH37422)}, 
	title={The real-time motion control core of the Orocos project}, 
	year={2003},
	volume={2},
	number={},
	publisher = {IEEE},
	address ={Taipei, Taiwan},
	pages={2766-2771 vol.2},
	doi={10.1109/ROBOT.2003.1242011}
}

@inproceedings{10.1145/3368089.3409743,
	author = {Garc\'{\i}a, Sergio and Str\"{u}ber, Daniel and Brugali, Davide and Berger, Thorsten and Pelliccione, Patrizio},
	title = {Robotics software engineering: a perspective from the service robotics domain},
	year = {2020},
	isbn = {9781450370431},
	publisher = {Association for Computing Machinery},
	address = {New York, NY, USA},
	url = {https://doi.org/10.1145/3368089.3409743},
	doi = {10.1145/3368089.3409743},
	booktitle = {Proceedings of the 28th ACM Joint Meeting on European Software Engineering Conference and Symposium on the Foundations of Software Engineering},
	pages = {593–604},
	numpages = {12},
	location = {Virtual Event, USA},
	series = {ESEC/FSE 2020}
}

@inproceedings{10.1145/3426425.3426942,
	author = {Ghzouli, Razan and Berger, Thorsten and Johnsen, Einar Broch and Dragule, Swaib and W\k{a}sowski, Andrzej},
	title = {Behavior trees in action: a study of robotics applications},
	year = {2020},
	isbn = {9781450381765},
	publisher = {Association for Computing Machinery},
	address = {New York, NY, USA},
	url = {https://doi.org/10.1145/3426425.3426942},
	doi = {10.1145/3426425.3426942},
	booktitle = {Proceedings of the 13th ACM SIGPLAN International Conference on Software Language Engineering},
	pages = {196–209},
	numpages = {14},
	location = {Virtual, USA},
	series = {SLE 2020}
}

@book{glad2018control,
	title={Control theory},
	author={Glad, Torkel and Ljung, Lennart},
	year={2000},
	publisher={CRC press},
	address={Boca Raton, Florida},
	mumpages={482},
	doi={https://doi.org/10.1201/9781315274737}
}

@article{trentelman2002control,
	title={Control theory for linear systems},
	author={Trentelman, Harry L and Stoorvogel, Anton A and Hautus, Malo and Dewell, L},
	journal={Appl. Mech. Rev.},
	volume={55},
	number={5},
	pages={B87--B87},
	year={2002}
}

@article{zheng2014state,
	title={On the state of the art in verification and validation in cyber physical systems},
	author={Zheng, Xi and Julien, Christine and Kim, Miryung and Khurshid, Sarfraz},
	journal={The University of Texas at Austin, The Center for Advanced Research in Software Engineering, Tech. Rep. TR-ARiSE-2014-001},
	volume={1485},
	year={2014},
	publisher={Citeseer},
	numpages={12},
}

@article{mandrioli2023stress,
	title={Stress Testing Control Loops in Cyber-Physical Systems},
	author={Mandrioli, Claudio and Shin, Seung Yeob and Maggio, Martina and Bianculli, Domenico and Briand, Lionel},
	journal={ACM Transactions on Software Engineering and Methodology},
	volume={33},
	number={2},
	pages={1--58},
	year={2023},
	publisher={ACM New York, NY}
}

@article{krichen2023survey,
	title={A survey on formal verification and validation techniques for internet of things},
	author={Krichen, Moez},
	journal={Applied Sciences},
	volume={13},
	number={14},
	pages={8122},
	year={2023},
	publisher={MDPI}
}

@ARTICLE{704994,
	author={Scokaert, Pierre O.M. and Rawlings, James B. },
	journal={IEEE Transactions on Automatic Control}, 
	title={Constrained linear quadratic regulation}, 
	year={1998},
	volume={43},
	number={8},
	pages={1163-1169},
	doi={10.1109/9.704994}
}

@article{GARCIA1989335,
	title = {Model predictive control: Theory and practice—A survey},
	journal = {Automatica},
	volume = {25},
	number = {3},
	pages = {335-348},
	year = {1989},
	issn = {0005-1098},
	doi = {https://doi.org/10.1016/0005-1098(89)90002-2},
	url = {https://www.sciencedirect.com/science/article/pii/0005109889900022},
	author = {Carlos E. García and David M. Prett and Manfred Morari}
}

@ARTICLE{8792042,
	author={Lee, Kibeom and Jeon, Seungmin and Kim, Heegwon and Kum, Dongsuk},
	journal={IEEE Access}, 
	title={Optimal Path Tracking Control of Autonomous Vehicle: Adaptive Full-State Linear Quadratic Gaussian (LQG) Control}, 
	year={2019},
	volume={7},
	number={},
	pages={109120-109133},
	doi={10.1109/ACCESS.2019.2933895}
}

@book{bubnicki2005modern,
	title={Modern control theory},
	author={Bubnicki, Zdzislaw and others},
	volume={2005925392},
	year={2005},
	publisher={Springer},
	address={Berlin Heidelberg},
	doi={https://doi.org/10.1007/3-540-28087-1}
}

@article{YU201546,
	title = {A survey of fault-tolerant controllers based on safety-related issues},
	journal = {Annual Reviews in Control},
	volume = {39},
	pages = {46-57},
	year = {2015},
	issn = {1367-5788},
	doi = {https://doi.org/10.1016/j.arcontrol.2015.03.004},
	url = {https://www.sciencedirect.com/science/article/pii/S136757881500005X},
	author = {Xiang Yu and Jin Jiang}
}

@ARTICLE{1220756,
	author={Ming-Tzu Ho and Chia-Yi Lin},
	journal={IEEE Transactions on Automatic Control}, 
	title={PID controller design for robust performance}, 
	year={2003},
	volume={48},
	number={8},
	pages={1404-1409},
	doi={10.1109/TAC.2003.815028}
}

@article{GARPINGER2014568,
	title = {Performance and robustness trade-offs in PID control},
	journal = {Journal of Process Control},
	volume = {24},
	number = {5},
	pages = {568-577},
	year = {2014},
	issn = {0959-1524},
	doi = {https://doi.org/10.1016/j.jprocont.2014.02.020},
	url = {https://www.sciencedirect.com/science/article/pii/S0959152414000730},
	author = {Olof Garpinger and Tore Hägglund and Karl Johan Åström}
}

@article{malu2014kinematics,
	title={Kinematics, localization and control of differential drive mobile robot},
	author={Malu, Sandeep Kumar and Majumdar, Jharna and others},
	journal={Global Journal of Research In Engineering},
	volume={14},
	number={1},
	pages={1--9},
	year={2014}
}

@ARTICLE{937419,
	author={Yongoug Chung and Chongkug Park and Harashima, F.},
	journal={IEEE Transactions on Industrial Electronics}, 
	title={A position control differential drive wheeled mobile robot}, 
	year={2001},
	volume={48},
	number={4},
	pages={853-863},
	doi={10.1109/41.937419}
}

@article{higuchi1993optimal,
	title={Optimal control of four wheel steering vehicle},
	author={Higuchi, Akira and Saitoh, Yasushi},
	journal={Vehicle System Dynamics},
	volume={22},
	number={5-6},
	pages={397--410},
	year={1993},
	publisher={Taylor \& Francis}
}

@article{SongOmni,
	author = {Song, Jae-Bok and Byun, Kyung-Seok},
	title = {Design and Control of a Four-Wheeled Omnidirectional Mobile Robot with Steerable Omnidirectional Wheels},
	journal = {Journal of Robotic Systems},
	volume = {21},
	number = {4},
	pages = {193-208},
	doi = {https://doi.org/10.1002/rob.20009},
	year = {2004}
}

@INPROCEEDINGS{9599335,
	author={Caldas, Ricardo and Ghzouli, Razan and Papadopoulos, Alessandro V. and Pelliccione, Patrizio and Weyns, Danny and Berger, Thorsten},
	booktitle={2021 IEEE International Conference on Autonomic Computing and Self-Organizing Systems Companion (ACSOS-C)}, 
	title={Towards Mapping Control Theory and Software Engineering Properties using Specification Patterns}, 
	year={2021},
	volume={},
	number={},
	pages={281-286},
	publisher={IEEE},
	address={DC, USA},
	doi={10.1109/ACSOS-C52956.2021.00067}
}

@inproceedings{10.1145/3387939.3391568,
	author = {C\'{a}mara, Javier and Papadopoulos, Alessandro V. and Vogel, Thomas and Weyns, Danny and Garlan, David and Huang, Shihong and Tei, Kenji},
	title = {Towards bridging the gap between control and self-adaptive system properties},
	year = {2020},
	isbn = {9781450379625},
	publisher = {Association for Computing Machinery},
	address = {New York, NY, USA},
	url = {https://doi.org/10.1145/3387939.3391568},
	doi = {10.1145/3387939.3391568},
	booktitle = {Proceedings of the IEEE/ACM 15th International Symposium on Software Engineering for Adaptive and Self-Managing Systems},
	pages = {78–84},
	numpages = {7},
	location = {Seoul, Republic of Korea},
	series = {SEAMS '20}
}

@INPROCEEDINGS{7194659,
	author={Filieri, Antonio and Maggio, Martina and Angelopoulos, Konstantinos and D'Ippolito, Nicolas and Gerostathopoulos, Ilias and Hempel, Andreas Berndt and Hoffmann, Henry and Jamshidi, Pooyan and Kalyvianaki, Evangelia and Klein, Cristian and Krikava, Filip and Misailovic, Sasa and Papadopoulos, Alessandro Vittorio and Ray, Suprio and Sharifloo, Amir M. and Shevtsov, Stepan and Ujma, Mateusz and Vogel, Thomas},
	booktitle={2015 IEEE/ACM 10th International Symposium on Software Engineering for Adaptive and Self-Managing Systems}, 
	title={Software Engineering Meets Control Theory}, 
	year={2015},
	volume={},
	number={},
	pages={71-82},
	publisher={IEEE},
	address={Florence, Italy},
	doi={10.1109/SEAMS.2015.12}
}

@ARTICLE{6313163,
	author={Luh, Johnson Yang-Seng},
	journal={IEEE Transactions on Systems, Man, and Cybernetics}, 
	title={Conventional controller design for industrial robots — A tutorial}, 
	year={1983},
	volume={SMC-13},
	number={3},
	pages={298-316},
	doi={10.1109/TSMC.1983.6313163}
}

@article{rivera1986internal,
	title={Internal model control: PID controller design},
	author={Rivera, Daniel E and Morari, Manfred and Skogestad, Sigurd},
	journal={Industrial \& engineering chemistry process design and development},
	volume={25},
	number={1},
	pages={252--265},
	year={1986},
	publisher={ACS Publications}
}

@article{GUIOCHET201743,
	title = {Safety-critical advanced robots: A survey},
	journal = {Robotics and Autonomous Systems},
	volume = {94},
	pages = {43-52},
	year = {2017},
	issn = {0921-8890},
	doi = {https://doi.org/10.1016/j.robot.2017.04.004},
	url = {https://www.sciencedirect.com/science/article/pii/S0921889016300768},
	author = {Jérémie Guiochet and Mathilde Machin and Hélène Waeselynck}
}

@misc{Ralph21_empirical_standards,
      title={Empirical Standards for Software Engineering Research}, 
      author={Paul Ralph and Nauman bin Ali and Sebastian Baltes and Domenico Bianculli and Jessica Diaz and Yvonne Dittrich and Neil Ernst and Michael Felderer and Robert Feldt and Antonio Filieri and Breno Bernard Nicolau de França and Carlo Alberto Furia and Greg Gay and Nicolas Gold and Daniel Graziotin and Pinjia He and Rashina Hoda and Natalia Juristo and Barbara Kitchenham and Valentina Lenarduzzi and Jorge Martínez and Jorge Melegati and Daniel Mendez and Tim Menzies and Jefferson Molleri and Dietmar Pfahl and Romain Robbes and Daniel Russo and Nyyti Saarimäki and Federica Sarro and Davide Taibi and Janet Siegmund and Diomidis Spinellis and Miroslaw Staron and Klaas Stol and Margaret-Anne Storey and Davide Taibi and Damian Tamburri and Marco Torchiano and Christoph Treude and Burak Turhan and Xiaofeng Wang and Sira Vegas},
      year={2021},
      eprint={2010.03525},
      archivePrefix={arXiv},
      primaryClass={cs.SE}
}

@article{ALBONICO2023111574,
	title = {Software engineering research on the Robot Operating System: A systematic mapping study},
	journal = {Journal of Systems and Software},
	volume = {197},
	pages = {111574},
	year = {2023},
	issn = {0164-1212},
	doi = {https://doi.org/10.1016/j.jss.2022.111574},
	url = {https://www.sciencedirect.com/science/article/pii/S0164121222002503},
	author = {Michel Albonico and Milica Đorđević and Engel Hamer and Ivano Malavolta}
}

@misc{SEatGoogle,
	title={Software engineering at Google},
	author={Winters, Titus},
	isbn = {978-1492082798},
	publisher = {O'Reilly Media},
	year={2020},
	url={https://abseil.io/resources/swe-book},
}

@inbook{Wiley,	
	publisher = {John Wiley \& Sons, Ltd},
	author={Joseph L. Hellerstein and Yixin Diao and Sujay Parekh and Dawn M. Tilbury},
	title = {Feedback Control of Computing Systems},
	chapter = {Z-Transforms and Transfer Functions},
	pages = {65--109},
	year = {2004}
}

@inproceedings{10.5555/3283535.3283545,
	author = {Abbas, Houssam and Saha, Indranil and Shoukry, Yasser and Ehlers, R\"{u}diger and Fainekos, Georgios and Gupta, Rajesh and Majumdar, Rupak and Ulus, Dogan},
	title = {Embedded software for robotics: challenges and future directions: special session},
	year = {2018},
	isbn = {9781538655641},
	publisher = {IEEE Press},
	booktitle = {Proceedings of the International Conference on Embedded Software},
	articleno = {10},
	numpages = {10},
	location = {Turin, Italy},
	series = {EMSOFT '18}
}

@Article{Li2024,
	author={Li, Wei	and Ribeiro, Pedro and Miyazawa, Alvaro and Redpath, Richard and Cavalcanti, Ana and Alden, Kieran and Woodcock, Jim and Timmis, Jon},
	title={Formal design, verification and implementation of robotic controller software via RoboChart and RoboTool},
	journal={Autonomous Robots},
	year={2024},
	month={Jul},
	day={05},
	volume={48},
	number={6},
	pages={14},
	issn={1573-7527},
	doi={10.1007/s10514-024-10163-7},
	url={https://doi.org/10.1007/s10514-024-10163-7}
}

@book{weyns2020introduction,
	title={An introduction to self-adaptive systems: A contemporary software engineering perspective},
	author={Weyns, Danny},
	year={2020},
	publisher={John Wiley \& Sons}
}

@software{ros,
	author = {{Stanford Artificial Intelligence Laboratory}},
	title = {Robotic Operating System},
	url = {https://www.ros.org},
	version = {ROS Melodic Morenia},
	date = {2018-05-23},
}

@article{doi:10.1126/scirobotics.abm6074,
	author = {Steven Macenski and Tully Foote and Brian Gerkey and Chris Lalancette and William Woodall},
	title = {Robot Operating System 2: Design, architecture, and uses in the wild},
	journal = {Science Robotics},
	volume = {7},
	number = {66},
	pages = {eabm6074},
	year = {2022},
	doi = {10.1126/scirobotics.abm6074},
	URL = {https://www.science.org/doi/abs/10.1126/scirobotics.abm6074}
}

@inproceedings{10.1145/3195836.3195853,
	author = {Alami, Adam and Dittrich, Yvonne and W\k{a}sowski, Andrzej},
	title = {Influencers of quality assurance in an open source community},
	year = {2018},
	isbn = {9781450357258},
	publisher = {Association for Computing Machinery},
	address = {New York, NY, USA},
	url = {https://doi.org/10.1145/3195836.3195853},
	doi = {10.1145/3195836.3195853},
	booktitle = {Proceedings of the 11th International Workshop on Cooperative and Human Aspects of Software Engineering},
	pages = {61–68},
	numpages = {8},
	location = {Gothenburg, Sweden},
	series = {CHASE '18}
}

@INPROCEEDINGS{6413422,
	author={Khan, Muhammad Fayaz and ul Islam, Raza and Iqbal, Jamshed},
	booktitle={2012 International Conference of Robotics and Artificial Intelligence}, 
	title={Control strategies for robotic manipulators}, 
	year={2012},
	volume={},
	number={},
	pages={26-33},
	doi={10.1109/ICRAI.2012.6413422}
}

@article{schwenzer2021review,
	title={Review on model predictive control: An engineering perspective},
	author={Schwenzer, Max and Ay, Muzaffer and Bergs, Thomas and Abel, Dirk},
	journal={The International Journal of Advanced Manufacturing Technology},
	volume={117},
	number={5},
	pages={1327--1349},
	year={2021},
	publisher={Springer},
	doi={10.1007/s00170-021-07682-3},
	url={https://doi.org/10.1007/s00170-021-07682-3}
}

@article{Metta:YARP:2006,
	author = {Metta, Giorgio and Fitzpatrick, Paul and Natale, Lorenzo},
	journal = {International Journal of Advanced Robotics Systems, special issue on Software Development and Integration in Robotics},
	number = 1,
	title = {YARP: Yet Another Robot Platform},
	volume = 3,
	year = 2006
}

@article{pan2025survey,
	title={Survey on recent advances in planning and control for collaborative robotics},
	author={Pan, Ya-Jun and Buchanan, Scott and Chen, Qiguang and Wan, Lucas and Chen, Nuo and Forbrigger, Shane and Smith, Sean},
	journal={IEEJ Journal of Industry Applications},
	volume={14},
	number={2},
	pages={139--151},
	year={2025},
	publisher={The Institute of Electrical Engineers of Japan}
}

@inproceedings{quigley2009ros,
	title={ROS: an open-source Robot Operating System},
	author={Quigley, Morgan and Conley, Ken and Gerkey, Brian and Faust, Josh and Foote, Tully and Leibs, Jeremy and Wheeler, Rob and Ng, Andrew Y and others},
	booktitle={ICRA workshop on open source software},
	volume={3},
	number={3.2},
	pages={5},
	year={2009},
	organization={Kobe}
}

@book{ogata1995discrete,
  title={Discrete-time control systems},
  author={Ogata, Katsuhiko},
  year={1995},
  publisher={Prentice-Hall, Inc.}
}

@article{di2019sampled,
  title = {Sampled-data emulation of dynamic output feedback controllers for nonlinear time-delay systems},
  journal = {Automatica},
  year = {2019},
  doi = {https://doi.org/10.1016/j.automatica.2018.10.022},
  author = {Mario {Di Ferdinando} and Pierdomenico Pepe},
}

@article{cervin2003does,
  title={How does control timing affect performance? Analysis and simulation of timing using Jitterbug and TrueTime},
  author={Cervin, Anton and Henriksson, Dan and Lincoln, Bo and Eker, Johan and Arzen, K-E},
  journal={IEEE control systems magazine},
  year={2003},
  doi={10.1109/MCS.2003.1200240}
}

@article{cervin2002feedback,
  title={Feedback--feedforward scheduling of control tasks},
  author={Cervin, Anton and Eker, Johan and Bernhardsson, Bo and {\AA}rz{\'e}n, Karl-Erik},
  journal={Real-Time Systems},
  volume={23},
  number={1},
  pages={25--53},
  year={2002},
  publisher={Springer}
}

@book{buttazzo2024hard,
  title={Hard real-time computing systems: predictable scheduling algorithms and applications},
  author={Buttazzo, Giorgio C},
  year={2024},
  publisher={Springer},
  doi = {10.1007/978-3-031-45410-3},
}


\end{document}